\documentstyle[amssymb,preprint,aps]{revtex}

\begin{document}
\title{From inversion to enhancement of the Josephson current by an exchange field
in S/F-I-F/S tunnel structures}
\author{V. N. Krivoruchko, E.A.Koshina}
\address{Donetsk Physics \& Technology Institute, Donetsk , Ukraine}
\date{\today}
\maketitle

\begin{abstract}
We study theoretically the dc Josephson tunnel current for a junction of two
proximity S/F bilayers of a massive superconductor (S) / a thin ferromagnet
(F) separated by an insulating (I) barrier. The dependence of the critical
current on the relative orientation of the F layers magnetization is
analyzed within the microscopic theory of the proximity effect for an S/F
bilayer. We demonstrate that for the S/F-I-F/S contact critical current can
reverse its sign ($\pi $-state of the junction) for the parallel orientation
of the F layers magnetization, wile for antiparallel alignment an
enhancement of the critical current takes place. The results provide a new
effect of the interplay between superconductivity and ferromagnetism in
hybrid structures.

\end{abstract}

\bigskip

Progress in nanotechnology in the last few years made it possible to produce
nanostructures in which new physical phenomena have been observed.
Specifically, hybrid systems consisting of superconductors (S) and
ferromagnets (F) have been created, which opens a possibility to explore
various mesoscopic effects in their superconducting and magnetic properties.
Particularly, the transport properties of S/F structures with artificial
geometry have turned out to be quite unusual. E.g., for SFS weak links a
crossover from $0$-phase to $\pi $-phase superconductivity function of the
thickness d$_{F}$ of ferromagnet,\ or the exchange field $H_{exc}$, have
been theoretically described [1,2] and experimentally observed [3], as well
as oscillation of the S/F multilayers transition temperatures [4-7].
Recently {\it Bergeret et al.} [8] predicted that in the case of an
antiparallel orientation of the exchange field of the bilayers, critical
current of the S/F-I-F/S junction increases at low temperatures with
increasing $H_{exc}$ and at zero temperature has a singularity when $H_{exc}$
equals the superconducting gap. This behavior contrasts common knowledge
that exchange field reduces the Josephson critical current. The authors
consider the model, when the influence of the F layers on superconductivity
is equivalent to inclusion of a homogeneous exchange field with a reduced
value, and come to limit of effective values of the superconducting order
parameter, the coupling constant, and the effective magnetic moment.
However, as is well known, any quantitative calculation of the properties of
inhomogeneous superconductors in contact with magnetic interfaces must start
with an accurate boundary conditions and calculation of the superconducting
properties near the S/F interface [9,10]. In real systems the exchange field
of the F layer leads to such decrease of the interface transparency that
allows a jump of the pairing amplitude near the interface, while theory [8]
assumes that the pairing function is continuous across the interface. As a
matter of fact, the results of Ref.8 should not be considered as conclusive
ones and it is reasonable to analyze a generalization of the model to a more
realistic case. In the present report, on the basis of microscopic theory of
the proximity effect for an S/F bilayer [11], the critical current of an
S/F-I-F/S tunnel junction is analyzed as a function of the parameters of the
S/F bilayer, such as the proximity-effect magnitude, the transparency of the
S/F interface, the exchange-field strength, and relative magnetizations
orientation of the ferromagnetic layers.

We consider the bilayer of a massive superconductor and a thin ferromagnet,
where (singlet) superconducting and ferromagnetic metals are assumed to be
dirty. It is shown that by changing the relative magnetizations orientation
one can for the same symmetric\ S/F-I-F/S contact turn the tunnel current
from $0$-phase state to $\pi $-phase state (inversion of the critical
current), if there is parallel orientation of the F layers magnetization, or
enhance the tunnel current, if there is antiparallel orientation of the F
layers.\ The feature important for practical application is that the
conditions for critical current inversion or enhancement can be changed by
varying the parameters of the S/F bilayer and external magnetic field
orientation. We also show that critical current singularity predicted in
Ref.8 is the result of the oversimplified model the authors considered.

To be specific, let us consider the proximity coupled S/F bilayer of a
massive superconductor and a thin ferromagnet, with the thickness of the F
layer much less than its superconducting coherence length $\xi _{F}\gg $d$%
_{F}$ ($\xi _{F}\symbol{126}\sqrt{\hbar D_{F}/\mu _{B}H_{exc}}$ where $D_{F}$%
\ is a diffusion coefficient in the F metal).\ One can expect a kind of
tunneling interaction in the S/F-sandwich that is similar to the
superconducting proximity effect. Actually, when diffusing into the thin
ferromagnetic layer, superconducting electrons are subject to an interaction
with the local exchange field. E.g., an electron with a spin, e.g. ''up'',
has an extra energy $\mu _{B}H_{exc}$ , while an electron with a spin
''down'' has lower energy $\mu _{B}H_{exc}$, caused by the intrinsic
magnetic field $H_{exc}$ into F layer (see, e.g. [12]). After tunneling back
to S layer the Cooper pair quickly loses its extra energy during the time $%
\tau \symbol{126}\hslash /\mu _{B}H_{exc}$ being related to a small range $%
\lambda _{F}\symbol{126}\hslash v_{F}/\mu _{B}H_{exc}$ ($v_{F}$ - Fermi
velocity). Such an equilibrium process leads to a modification of the
electron spectrum of superconductor on the nanoscale length $\lambda _{F}$
(in most cases $\lambda _{F}\symbol{126}10^{-8}m$). So, we can speak about
induced exchange magnetic correlation into the S layer that affects the
Cooper pairs and characterize the S/F bilayer as a unified system with
strong superconducting-ferromagnetic correlation. Owing to this magnetic
proximity effect the S/F-I-F/S junction can get the $\pi $-phase
superconductivity even if the thickness of the F layer is much less than the
superconducting coherence length of the F metal . This mechanism should not
be confused with the $\pi $-junction behavior induced by magnetic impurities
[13,14], or resulting from the symmetry of the order parameter [15], or due
to direct access to the microscopic current-carrying electronic state inside
the link [16,17]. Our case also differs from the situation in S/F sandwiches
with thick F layer, where spatially dependent phase in the F layer causes an
exchange field dependent oscillation in the critical current of the SFS weak
links and in the T$_{C}$ of the multilayers [1-7] (for details see Refs. 11,
18).

The critical current of the (S/F)$_{L}$-I-(F/S)$_{R}$ tunnel contact can be
represented in the form (Ref.18)

\begin{center}
$j_{C}=(eR_{N}/2\pi T_{C})I_{C}=(T/T_{C})%
\mathop{\rm Re}%
\sum_{\omega >0}\{G_{SL}\Phi _{SL}G_{SR}\Phi _{SR}/\omega ^{2}\}\times $ \ \
\ \ \ \ \ \ \ \ \ \ \ \ \ \ \ (1) $\{[1+2\varpi G_{S}(\gamma _{B}/\pi
T_{C})+\varpi ^{2}(\gamma _{B}/\pi T_{C})^{2}]_{L}\times \lbrack 1+2\varpi
G_{S}(\gamma _{B}/\pi T_{C})+\varpi ^{2}(\gamma _{B}/\pi
T_{C})^{2}]_{R}\}^{-1/2}$
\end{center}

where R$_{N}$ is the resistance of the contact in the normal state; $\varpi
=\omega +i(\pm H_{exc})$ and the sign of the exchange field depends on
mutual orientation of the bank magnetizations; $\omega =\pi T(2n+1)$ are
Matsubara frequencies; and the subscript L (R) labels quantities referring
to the left (right) bank. Hencerforth, we have taken the system of units in
with $\hbar =\mu _{B}=k_{B}=1$, and have also used the modified [19] Usadel
function $\Phi _{S,F}$ defined by relations $G_{S}=\omega /(\omega ^{2}+\Phi
_{S}^{2})^{1/2}$ , $\digamma _{S}=G_{S}\Phi _{S}/\omega $ , $G_{F}=\varpi
/(\varpi ^{2}+\Phi _{F}^{2})^{1/2}$ , $\digamma _{F}=G_{F}\Phi _{F}/\varpi $
. So, we have taken explicitly the normalization condition $G^{2}+|F|^{2}=1$
on usual Usadel functions $G_{S,F}$ , $F_{S,F}$ .

Presented below are the results obtained on the basis of microscopic theory
of proximity effect for an S/F bilayer characterized by arbitrary values of
the exchange field $H_{exc}$, S/F interface boundary transparency $\gamma
_{B}$ and for the cases of a weak ($\gamma _{M}<<1$) or a strong ($\gamma
_{M}>>1$) proximity effect (Ref.11). Namely, for a weak proximity effect we
have for the function $\Phi _{S}(\omega ):$

\begin{center}
$\ \ \ \ \ \ \ \ \ \ \ \ \ \ \ \ \ \ \ \ \ \ \ \ \ \ \ \ \ \ \ \ \ \Phi
_{S}(\omega )=\Delta _{0}\left( 1-%
{\displaystyle{\gamma _{M}\beta \varpi  \over \gamma _{M}\beta \varpi +\omega A}}%
\right) $ \ \ \ \ \ \ \ \ \ \ \ \ \ \ \ \ \ \ \ \ \ \ \ \ \ \ \ \ \ \ \ \ \
\ \ \ (2)
\end{center}

where $\Delta _{0}$ is the absolute value of the BCS order parameter in the
bulk of the S layer,

$\beta ^{2}=\left( \omega ^{2}+\Delta _{0}^{2}\right) ^{1/2}/\pi T_{C},$
and\ $A=\left[ 1+\gamma _{B}\varpi \left( \gamma _{B}\varpi +2\omega /\beta
^{2}\right) /(\pi T_{C})^{2}\right] ^{1/2}$. For a strong proximity effect
our calculations yield

\begin{center}
$\ \ \ \ \ \ \ \ \ \ \ \ \ \ \ \ \ \ \ \ \ \ \ \ \ \ \ \ \ \ \ \ \ \ \ \ \
\Phi _{S}(\omega )=B(T)\left( \pi T_{C}+\gamma _{B}\varpi \right) /\gamma
_{M}\varpi $ , \ \ \ \ \ \ \ \ \ \ \ \ \ \ \ \ \ \ \ \ \ \ \ \ \ \ \ \ \ \ \
\ \ \ \ (3)
\end{center}

where $B(T)=2T_{C}[1-(T/T_{C})^{2}][7\zeta (3)]^{-1/2}$ and $\zeta (3)$ is
the Riemann $\zeta $ function. As is seen from the expressions (2), (3)
ferromagnetic correlations are induced into S layer and Green's functions of
the S layer now depend upon $H_{exc}$. Using these results, we calculate the
dependence of the amplitude of S/F-I-F/S junction critical current on the
orientation of the magnetization in the F layers.

{\it Parallel orientation of the layer's magnetizations. }Let us present
here an analytical consideration for the case of a vanishing interface
resistance, $\gamma _{B}=0$. Using the expressions (1) and (2) with $\varpi
_{L}=\varpi _{R}$, we have obtained for a weak influence of the F layer on
superconducting properties of the S metal, $\gamma _{M}<<1$:

\begin{center}
$\ \ \ \ \ \ \ \ \ \ \ \ \ \ \ \ \ \ \ j_{C}^{p}$ $\ \symbol{126}$ $\Delta
_{0}^{2}\sum_{\omega >0}%
{\displaystyle{\Delta _{0}^{2}+\omega ^{2}-(\gamma _{M}\beta H_{exc})^{2} \over \left[ \Delta _{0}^{2}+\omega ^{2}-(\gamma _{M}\beta H_{exc})^{2}\right] ^{2}+(2\omega \gamma _{M}\beta H_{exc})^{2}}}%
$ \ \ \ \ \ \ \ \ \ \ \ \ \ \ \ \ \ (4)
\end{center}

One can see, that if the exchange field is strong enough, namely $\gamma
_{M}H_{exc}>\pi T_{C}\beta $, the critical current changes its sign; that is
the phase difference between the superconducting order parameters on banks
of the junction changes by $\pi $ .

If there is a strong suppression of the order parameter near SF boundary, $%
\gamma _{M}>>1$ , critical current of the contact can be presented in the
form:

\begin{center}
$\ \ \ \ \ \ \ \ \ \ \ \ \ \ \ \ \ \ j_{C}^{p}\approx
(T/T_{C})B_{M}^{2}(T)2\sum_{\omega }(\omega ^{2}-H_{exc}^{2})/\{\omega
^{2}(\omega ^{2}+H_{exc}^{2})^{2}\}$ \ \ \ \ \ \ \ \ \ \ \ \ \ \ \ \ \ \ \ \
(5)
\end{center}

where $B_{M}(T)=B(T)\pi T_{C}/\gamma _{M}$\ , and in proceeding these
relations, we have taken into account that value of $\Phi _{S}$ is small, $%
\Phi _{S}\symbol{126}\gamma _{M}^{-1}$. For $H_{exc\text{ }}\rightarrow 0$
expressions (4) and (5) restore the result for S/N-I-N/S junction (see,
e.g., Ref.19). In the opposite case of an increasing magnetic energy the
critical current changes its sign for large enough\ $H_{exc\text{ }}$, $%
H_{exc}>>\pi T_{C}$, or, in other words, the crossover of the junction from
0-phase state to the $\pi $-phase state takes place.

{\it Antiparallel orientation of the layer's magnetizations. }To be
definite, we took $\varpi _{L}=\omega +iH_{exc}$, $\varpi _{R}=\omega
-iH_{exc}$. After simple transformations we have for the case $\gamma _{B}=0$
and $\gamma _{M}<<1$:

\begin{center}
$j_{C}^{a}=T/T_{C}%
\mathop{\rm Re}%
\sum\limits_{\omega >0}%
{\displaystyle{\Phi _{S} \over \sqrt{\omega ^{2}+\Phi _{S}^{2}}}}%
\mid _{L}%
{\displaystyle{\Phi _{S} \over \sqrt{\omega ^{2}+\Phi _{S}^{2}}}}%
\mid _{R}\approx $

$2T/T_{C}\sum_{\omega >0}%
{\displaystyle{\Delta _{0}^{2} \over \omega ^{2}+\Delta _{0}^{2}}}%
\left[ 1-2(\gamma _{M}\beta H_{exc})^{2}%
{\displaystyle{\Delta _{0}^{2}-\omega ^{2} \over (\omega ^{2}+\Delta _{0}^{2})^{2}}}%
+%
{\displaystyle{(\gamma _{M}\beta H_{exc})^{4} \over (\omega ^{2}+\Delta _{0}^{2})^{2}}}%
\right] ^{-1/2}$
\end{center}

In proceeding these relations, we have taken into account that $\gamma
_{M}\beta $ is small, $\gamma _{M}\beta <<1$. One can see that for $\omega
<\Delta _{0}$ the expression in square brackets is lower than 1 for $H_{exc}$%
\ from a broad region $0<H_{exc}<$ $\sqrt{2(\Delta _{0}^{2}-\omega ^{2})}%
/\gamma _{M}\beta $. As the\ result, for some values of $H_{exc}$ one can
obtain the enhancement of the tunnel current, $%
j_{C}^{a}(H_{exc})>j_{C}^{a}(0)$, in contrast to its suppression by the
magnetic moments aligned in parallel.

For the case of a strong proximity effect $\gamma _{M}>>1$ we obtain:

\begin{center}
$j_{C}^{a}$ $\symbol{126}$ $B_{M}^{2}(T)T/T_{C}\sum_{\omega >0}\omega ^{-2}%
\left[ (\omega ^{2}-H_{exc}^{2}+B_{M}^{2}(T)/\omega ^{2})^{2}+(2\omega
H_{exc})^{2}\right] ^{-1/2}$
\end{center}

Now the exchange field region where critical current increase is much less $%
0<H_{exc}\leq \sqrt{B_{M}^{2}(T)-\omega ^{4}}/\omega $\ , however at low
temperature there is also the region where $j_{C}^{a}(H_{exc})>j_{C}^{a}(0)$.

For a general configuration, when the magnetizations of the banks are at an
angle $\theta $ , the conductivity for parallel channel is proportional to $%
\cos ^{2}(\theta /2)$, while the conductivity for antiparallel channel is
proportional to $\sin ^{2}(\theta /2)$. So, the critical current can be
written in the form [21, 8]:

\begin{center}
$j(\theta )=j_{C}^{p}\cos ^{2}(\theta /2)+j_{C}^{a}\sin ^{2}(\theta /2)$
\end{center}

As it was before, we suppose that the Hamiltonian involved with the
tunneling process is spin independent.

On Figs. 1 and 2 we show the results of numerical calculations of the
amplitude of the Josephson current for the case of a weak, $\gamma _{M}<<1$
(Fig.1), and strong $\gamma _{M}>>1$ (Fig.2) proximity effect and different
quality of the S and F metals electrical contact versus the exchange field
strength for parallel (solid curves) and antiparallel (dashed curves) mutual
orientation of the electrodes' magnetizations. It can be seen, that a state
of the junction depends greatly on the bilayer parameters $H_{exc}$, $\gamma
_{M}$ and $\gamma _{B}$, and relative orientation of the left and right
magnetizations. For the case of parallel orientation (solid curves on Figs.
1, 2) the critical current drops down to zero and then acquires negative
values. So, in some interval of exchange field strengths, a state with $\pi $
phase shift across the contact is formed. At weak proximity effect and high
boundary transparency, varying the bilayer parameters, we can change sizably
the conditions under which the state characterized by a $\pi $ phase
difference at the banks of the junction is realized (see Fig.1). At strong
proximity effect there is a reduction of the absolute value of $j_{C}$ , and
the point of the crossover is not so sensitive to the S/F boundary
conditions (see Fig.2).

For the case of antiparallel orientation (dashed curves on Figs. 1 and 2) a
state with $0$ phase shift across the contact is formed, but in some
interval of exchange field strengths the enhancement of dc Josephson current
takes place. As in the case of parallel orientation, if $\gamma _{B}<<1$ and 
$\gamma _{M}<<1$ by varying the bilayer parameters we can change sizably the
range of exchange field where the current enhancement is observed. If $%
\gamma _{M}>>1$ the role of the interface is reduced. As is seen on Figs. 1
and 2, for antiparallel geometry there is not singularity predicted in Ref.8.

Figure 3 shows the dc tunnel current versus exchange field at different
temperature. One interesting point to note is that the Josephson current
enhancement holds only for low enough temperatures, while at $T>0.5T_{C}$
the phenomenon disappears - see results presented on Fig.3 by dashed curves.
On the contrary, a spontaneous $\pi $ shift of superconducting wave
functions phase of the banks holds for all the superconducting state
temperature range.

The physics behind both striking behaviors of dc Josephson current in
question is the induced magnetic properties of the S layers. Namely, the S
metal in good electric contact with the F one acquires some magnetic
properties and one can characterize the S/F bilayer as unified system with
strong superconducting--ferromagnetic correlation. As a result, the
superconducting order parameter at the F/S boundary acquires phase shift
depending on the orientation and value of the exchange field in the F layers
[18]. Due to the Cooper pair amplitude fluctuation as a function of $H_{exc}$%
, there is a possibility at some value of the exchange field to arrange the
minimum of the pair's amplitude exactly on S/F interface. As a result, the
relative S/F boundary influence on superconducting order will be even lower
than in a case of S/N boundary with the same interface parameters $\gamma
_{B}$ and $\gamma _{M}$.

The enhancement of the Josephson current by an exchange field in
superconductor was recently discussed by {\it Bergeret et. al}.(Ref.8)
considering a toy model. However,the effects of inversion at parallel and
enhancement at antiparallel configurations for S/F-I-F/S junction found in
our work have not been discussed yet. The results of our report are based on
microscopic theory of the proximity effect for S/F bilayer and accurate
calculation of the superconducting properties near the S/F interface. By
using general expressions, obtained in [18], the amplitude of the critical
current of symmetric S/F-I-F/S tunnel contact has been\ calculated as
function of the F layers magnetization orientation and the S/F interface
parameters such as the proximity-effect strength, the transparency of the
S/F inerface, and the strength of the exchange field in the F metal, and
temperature. Our results show that the superconducting properties of
S/F-I-F/S junctions based on S/F bilayers of a massive superconductor and a
thin ferromagnet can be varied from a state with $0$-phase superconductivity
with enhancement of Josephson critical current to a $\pi $-phase
superconductivity with reversal of tunnel current by simply changing the
relative orientation of the left and right banks magnetization. These
results provide new effects of the proximity coupled
superconductor/ferromagnet hybrid structures.

We are grateful to V.V.Ryazanov who has turned our attention to the work [8].

\begin{center}
\bigskip
\end{center}

\bigskip


\bigskip

\begin{references}
\bibitem{}  1. A. I. Buzdin, L. N. Bulaevskii, S. V. Panjukov. Letters to
JETP .v35, p.147 (1982).

2. A. I. Buzdin, M. Yu. Kuprijanov, Letters to JETP n.53, p.308 (1991).

3. V. V. Ryazanov, V. A. Oboznov, A. Yu. Rusanov, A. V. Veretennikov,
A.A.Golobov, J. Aarts. cond-mat/0008364.

4. Z. Radovic, M. Ledvij, L. Dobrosavljevic-Grujic, A. I. Buzdin, J. R.
Clem. Phys.Rev. B 44, 759 (1991).

5. L. Lazar, K. Westerholt, H. Zabel, L. R. Tagirov, Yu. V. Goryunov, N. N.
Garif'yanov, I. A. Garifullin. Phys. Rev. B 61 3711 (2000).

6. J.S.Jiang, D. Davidovic, D.H.Reich, C. L. Cheien. Phys.Rev. Lett. 74, 314
(1995).

7. M. G. Khusainov Yu. N. Proshin. Phys. Rev. B 22, R14283 (1997).

8. F. S. Bergeret, A. F. Volkov, K. E. Efetov. cond-mat/0102012.

9. J.Aarts, J.M.Geers, E.Bruck, A.A.Golobov, R.Coehorn. Phys.Rev. B 56,
p.2779 (1997).

10. L.Lazar, K.Westerhold, H.Zabel, L.R.Tagirov, Yu.V.Gorynov,
N.N.Gafif'yanov, I.A.Garifullin. Phys.Rev. B 61, p.3711 (2000).

11. E. A. Koshina, V. N. Krivoruchko. Low. Tem. Phys.v.26, 115 (2000);

12. E. A. Demler, G. B. Arnold, M. R. Beasley, Phys.Rev. B 55, 15174 (1997).

13. L. N. Bulaevskii, V. V. Kuzii, A. A. Sobyanin. Pis'ma Zh. Eksp. Teor.
Fiz. v.25, p.314 (1977).

14. A. V. Andreev, A. I. Buzdin, R. M. Osgood. Phys. Rev. B 43, 10124 (1991).

15. M. Siegrist, T. M. Rice. J. Phys. Soc. Jpn. v.61, p.4283 (1992).

16. A. F. Volkov. Phys. Rev. Lett. v.74, p.4730 (1995).

17. J.J.A.\ Baselmans, A.F.Morpurgo, B. J. van Wees,T. Klapwijk. Nature 397,
43 (1999).

18. E. A. Koshina, V. N. Krivoruchko. JETP Letters v.71, 123 (2000);
Phys.Rev. B \ 2001 (to be published)

19. A. A. Golubov, M. Yu. Kuprijanov, JETP 96, 1420 (1989).

20. M. Yu. Kupriyanov, V. F. Lukichev. Sov. J. Low Tem. Phys. v.8, 1045
(1982).

21. P. Raychaudhuri, K. Sheshadri, P. Taneja, S. Banyopadhyay, P. Ayyub, A.
K. Nigam. R. Pinto. Phys.Rev. B 59, 13919 (1999).
\end{references}
\end{document}